\documentclass[aps,prd,superscriptaddress,11pt,showpacs,notitlepage]{revtex4}
	\usepackage{graphicx}
	\usepackage{amsmath,amssymb,mathrsfs}
\usepackage{epsf}
\usepackage{epsfig}

\usepackage{color}

\usepackage{amsmath}

\usepackage{amssymb}

\numberwithin{equation}{section}

\usepackage{subfigure}

\newcommand{\be}{\begin{equation}}
\newcommand{\ee}{\end{equation}}
\newcommand{\bea}{\begin{eqnarray}}
\newcommand{\eea}{\end{eqnarray}}
\newcommand{\pa}{\partial}
\newcommand{\bb}{\bibitem}
 
\newcommand{\eqn}{\begin{eqnarray}}
\newcommand{\eqnx}{\end{eqnarray}}

\begin{document}
\title{The volume of a soliton}

\author{C. Adam}
\affiliation{Departamento de F\'isica de Part\'iculas, Universidad de Santiago de Compostela and Instituto Galego de F\'isica de Altas Enerxias (IGFAE) E-15782 Santiago de Compostela, Spain}
\author{M. Haberichter}
\affiliation{School of Mathematics, Statistics and Actuarial Science, University of Kent, Canterbury, CT2 7NF, UK}
\author{A. Wereszczynski}
\affiliation{Institute of Physics,  Jagiellonian University,
Lojasiewicza 11, Krak\'{o}w, Poland}

\begin{abstract}
There exists, in general, no unique definition of the size (volume, area, etc., depending on dimension) of a soliton. Here we demonstrate that the geometric volume (area etc.) of a soliton is singled out in the sense that it exactly coincides with the thermodynamical or continuum-mechanical volume. In addition, this volume may be defined uniquely for rather arbitrary solitons in arbitrary dimensions.
\end{abstract}
\maketitle 

\section{Introduction}
Solitons are finite-energy solutions of nonlinear field equations with properties which make them similar to particles in several aspects \cite{raja-book,man-sut-book}. They are ubiquitous in condensed matter systems (see, e.g., \cite{abrik}-\cite{ezawa}) and, in addition, find some applications in cosmology \cite{vilenkin}-\cite{battye} and nuclear and elementary particle physics \cite{skyrme}-\cite{brown-rho}. For concreteness and simplicity, we shall restrict the following discussion 
to relativistic field theories with a Poincare-invariant action in $d$ space plus one time dimension, although our considerations may probably be adapted without difficulties to non-relativistic field theories. Relativistic theories with solitons support both static soliton solutions and their boosted versions. Besides its energy, further basic characteristics of a static soliton are its position and its size. While the energy of a soliton resulting from an action principle in a Poincare-invariant theory is always unique and well-defined, it is not so obvious how to find well-motivated definitions of its position and size. In particular, in the present contribution we want to investigate the possibility of a unique definition of the size of a soliton.

Once an adequate definition for the position $\vec x_0$ of a soliton has been found, owing to translational invariance there will exist solitons with arbitrary positions and otherwise unchanged properties. For a given position, most of the energy density $\mathcal{E}$ of the soliton will be concentrated in a finite region about this position, whereas for large distances the energy density may either decay sufficiently fast (e.g., exponentially or algebraically) or even be exactly equal to zero. What, then, might be possible definitions of the soliton size? If the energy density is spherically symmetric and takes its maximum value at the position of the soliton, then one possible definition of its radius $R$ might be the distance $|\vec x -\vec x_0|$ from the soliton position $\vec x_0$ where the energy density has diminished by a certain factor. The size of the soliton would then be the generalized volume $V_d$ (area, volume, etc.) of the $d$-dimensional ball $\mathbb{D}^d$ with radius $R$ in $d$-dimensional space, e.g., $V_3=(4\pi/3)R^3$ in $d=3$. If, in addition to the energy density,  there exists a further density $\rho$ interpretable as a particle density or charge density, then further possible radius definitions are the corresponding root-mean-square (RMS) radii and generalizations thereof, i.e., 
\be
R_\alpha \equiv \left( \frac{1}{A_{d-1}}\int d^d x |\vec x - \vec x_0|^\alpha \rho (\vec x) \right)^\frac{1}{\alpha}
\ee
(where $A_d$ is the generalized area of the $d$-dimensional unit sphere, e.g., $A_0 = 2$, $A_1 = 2\pi$, $A_2 = 4\pi$, etc.) Here, $\alpha =2$ corresponds to the RMS radius. In particular, this definition may be used for field theories possessing so-called topological solitons \cite{man-sut-book}, i.e., soliton solutions which obey some non-trivial boundary conditions. They are usually characterized by a (typically integer-valued) topological degree 
\be
Q = \int d^d x \rho_q (\vec x)
\ee
where the topological charge density $\rho_q$ (a function of the fields $\vec \phi$ of the theory and their first derivatives) may be expressed as a total derivative such that $Q$ only depends on the boundary conditions imposed on $\vec \phi$ and is invariant under local variations of the fields. $\rho_q$ may obviously be used to define (generalized) RMS radii. All these definitions of the soliton size, however, depend on an identification of its position, among other peculiarities.

\section{A two-dimensional example}
Let us consider a simple example in one space dimension. The standard lagrangian density for a real scalar field $\phi$ is (our Minkowski metric convention is $g_{00}=+1$)
\be \label{standard-lag}
{\cal L} = \frac{1}{2}\pa_\mu \phi \pa^\mu \phi - m^2 U(\phi) 
\ee
where $U$ is a non-negative potential. Further, we assume natural units and a dimensionless scalar field, such that both $\partial_\mu$ and $m$ have the dimension of mass. 
The energy momentum tensor is of the perfect fluid type and reads
\be
T_{\mu\nu} =  \pa_\mu \phi \pa_\nu \phi - g_{\mu\nu} {\cal L} = u_\mu u_\nu (\mathcal{E} + \mathcal{P}) - g_{\mu\nu} \mathcal{P}
\ee
where (assuming $\pa_\mu \phi \pa^\mu \phi <0$)
\be
u_\mu = \left( -\pa_\lambda \phi \pa^\lambda \phi \right)^{-\frac{1}{2}}\epsilon_{\mu\nu} \pa^\nu \phi
\ee
is the ``two-velocity" (the two-dimensional version of the four-velocity), and 
\bea 
 \mathcal{E} &=& -\frac{1}{2} \pa_\mu \phi \pa^\mu \phi + m^2 U \\
 \mathcal{P} &=&  -\frac{1}{2} \pa_\mu \phi \pa^\mu \phi - m^2 U
\eea
are the proper energy density and the (local) pressure density, respectively. We remark that, for non-static configurations, $\mathcal{E} \not= T_{00}$, $\mathcal{P} \not= T_{11}$ because $\mathcal{E}$ and $\mathcal{P}$ are Lorentz scalars. In the static case, we have ($\phi ' \equiv \pa_x \phi$, $\dot\phi \equiv \pa_t \phi$)
\bea 
T_{00} = \mathcal{E} &=& \frac{1}{2} \phi'^2 + m^2 U \\
T_{11} =  \mathcal{P} &=&  \frac{1}{2}  \phi'^2 - m^2 U.
\eea
Further,
the Euler--Lagrange equation reads
\be \label{eul-lag}
\pa_\mu \pa^\mu \phi  +m^2 U_{,\phi} =0 
\ee
which in the static case may be integrated once to 
\be \label{const-P-eq}
\mathcal{P} \equiv \frac{1}{2}\phi'^2 - m^2 U =P=\mbox{const}
\ee
where the pressure $P$ is the constant on-shell value of the pressure density $\mathcal{P}$. The same conclusion $\mathcal{P}=P=\mbox{const}$ follows, in fact, directly from energy-momentum conservation for static fields, $\partial_x T_{11} =\partial_x \mathcal{P}=0$. 

\subsection{Kinks}

Finite energy configurations can only exist if the potential $U(\phi)$ has at least one vacuum, i.e., a field value $\phi = \phi_1$ such that $U(\phi_1)=0$. If the potential has two vacua $\phi_1$, $\phi_2$, then there exist so-called kink solutions $\phi_{\rm k} (x)$, i.e., static solutions interpolating between the two vacua, e.g., $\lim_{x \to -\infty} \phi_{\rm k} = \phi_1$,   $\lim_{x \to +\infty} \phi_{\rm k} = \phi_2$. Configurations with these boundary conditions may be distinguished from configurations of the vacuum sector (approaching the same limit in both directions) by the (rather trivial) topological charge density 
\be
\rho_q = \frac{\phi '}{\phi_2 - \phi_1}
\ee
and the corresponding charge
\be
Q = \int_{-\infty}^\infty dx \rho_q =\int_{\phi_-}^{\phi_+} \frac{d\phi}{\phi_2 - \phi_1} =\frac{\phi_+ - \phi_-}{\phi_2 -\phi_1}
\ee
where $\phi_\pm \equiv \lim_{x\to \pm \infty} \phi(x)$ and both $\phi_+$ and $\phi_-$ must take any of the vacuum values $\phi_1$ and $\phi_2$. So $Q$ may take the values $0$ (vacuum sector), 1 (kink sector) and $-1$ (antikink sector). Configurations from different sectors cannot be transformed into each other by local (finite energy) deformations.  The energy of a kink (antikink) configuration is bounded from below by a similar topological bound (i.e., a bound only depending on the theory - the potential - and the boundary conditions)
\be
E= \frac{1}{2} \int_{-\infty}^\infty dx \left(  \phi '^2 + 2 m^2 U\right) = \frac{1}{2} \int dx  (\phi ' \mp m \sqrt{2U})^2 \pm  m\int dx \sqrt{2U} \phi '
\ee
i.e., 
\be
E \ge \left|  m\int dx \sqrt{2U} \phi ' \right| = m \left| \int_{\phi_-}^{\phi_+}d\phi \sqrt{2U} \right| = m|(W(\phi_+) - W(\phi_-)|
\ee
where the ``superpotential" $W(\phi)$ is defined via $W_{,\phi} = \sqrt{2U}$ and may be calculated easily for particular potentials $U$. The energy bound is saturated for solutions of the first-order (Bogomolny) equations 
\be
\phi '  =\pm m \sqrt{2U}.
\ee
These two equations are easily recognized as the two roots of the zero pressure equation, i.e., of the once-integrated static field equation (\ref{const-P-eq}) for zero pressure $P=0$ (this relation between the zero pressure condition and BPS equations has been pointed out already, e.g., in \cite{bazeia2}). 
Solutions of these equations are called kinks (plus sign) and antikinks (minus sign), respectively. They provide global energy minima in their respective topological sectors.

All above results remain valid for potentials with more than two vacua provided that $\phi_1$ and $\phi_2$ refer to two adjacent vacua (the total number of inequivalent topological sectors, of course, increases in this case).  

\subsection{The $\phi^4$ kink}

As one simple, well-known example, let us consider the so-called $\phi^4$ potential
\be \label{phi4-pot}
U_{\phi^4} =\frac{1}{2}(1-\phi^2)^2
\ee
with two vacua at $\phi_1 =-1$, $\phi_2 =1$, and
leading to the kink energy $E=m\int_{-1}^1 d\phi (1-\phi^2) =(4/3)m$. The Bogomolny equation for the kink is $\phi ' = m(1-\phi^2)$ or
\be
dx = m^{-1} \frac{d\phi}{1-\phi^2}
\ee
with the solution
\be
x-x_0 = m^{-1} \mbox{artanh}\; \phi \quad \Rightarrow \quad \phi_{\rm k} = \tanh m(x-x_0) 
\ee
where $x_0$ is an integration constant.
The topological charge density and energy density of the kink are
\be
\rho_q = \frac{1}{2}\phi'_k = \frac{m}{2} \cosh^{-2} m(x-x_0) 
\ee
\be
\mathcal{E} = {\phi'_k} ^2 = m^2 \cosh^{-4} m(x-x_0) .
\ee
In this simple model, it is quite obvious that the point $x=x_0$ should be identified with the kink position. First of all, both $\rho_q$ and $\mathcal{E}$ have their maxima at $x=x_0$. Further, the point $x=x_0$ in physical space corresponds to the point $\phi =0$ in field space which is singled out, as well. The potential (\ref{phi4-pot}) is reflection symmetric about $\phi =0$, and $\phi =0$ is the point halfway between the two vacua. In addition, the potential has a local maximum at $\phi =0$ (which explains why the energy density maximum is located there). Accepting the value $x_0$ as the kink position, we may then calculate, e.g., its RMS radius via
\be
R_2^2 = \frac{1}{2} \int dx (x-x_0)^2 \rho_q = \frac{1}{4}\int_{-1}^1 d\phi \; m^{-2} \mbox{artanh}^2 \, \phi = \frac{\pi^2}{24 m^2},
\ee
leading to the kink RMS size (length) $L=2R_2 = \frac{\pi}{\sqrt{6}m}$. 

We want to emphasize, however, that this result for the ``size" of the kink is based on several assumptions, each of which is to a certain degree arbitrary. Indeed, we could use one of the generalized RMS radii $R_\alpha$ instead of the standard RMS radius. We could also use the normalized energy density instead of the topological density in the definition of $R_\alpha$. Another source of arbitrariness is related to the necessity to identify the position of the soliton. In the case at hand, all reasonable definitions agree in that the position $x_0$ of the kink should be where the kink field takes the midpoint value $\phi_{\rm k} (x_0)=(1/2)(\phi_2 + \phi_1)=0$. This is related to the particularly simple form of the potential. For more complicated potentials without reflection symmetry, there exist several possible definitions for the kink position which will lead to different results, in general. The most obvious ones are: a) the position of the midpoint $\phi_{\rm k} (x_0) =(1/2)(\phi_2 +\phi_1)$; b) the position of the maximum of $\mathcal{E}$ or $\rho_q$ (the two coincide for the scalar field models considered here); c) the position $x_0$ where one-half of the kink energy is to the left and one-half to the right; d) the position where one-half of the topological charge is to the left and one-half to the right.

In higher dimensions, there may be some further complications in the identification of the soliton position. In our one-dimensional example, we could identify the soliton position with the pre-image of the ``anti-vacuum" (the midpoint $(1/2)(\phi_2 + \phi_1)$), because the number of dimensions of field space and physical space coincide. But this is not always the case. If, e.g., the dimension of physical space is higher, then even if an  anti-vacuum may be defined in field space, its pre-image in physical space will be a (one-, two- or higher-dimensional) submanifold of physical space $\mathbb{R}^d$ rather than a point. In these cases, finding a sensible definition of the position $\vec x_0$ of a soliton is even more difficult and unavoidably plagued by a certain amount of arbitrariness. 

These considerations make one wonder whether there exists a more unequivocal definition of the size of a soliton. We shall see that this is indeed the case, although the resulting size (generalized volume) has some counterintuitive properties. 

\subsection{Kink size and non-zero pressure} 

Let us consider a general non-negative potential with two adjacent vacua $\phi_1$ and $\phi_2$ (and $\phi_2 >\phi_1$, without loss of generality). The two roots of the constant pressure equation
(\ref{const-P-eq}) read
\be \label{kink-eq}
\phi' = \pm \sqrt{2}\sqrt{m^2 U + P}
\ee
where we only consider the plus sign (kink) in the following. Separation of variables leads to
\be \label{sep-var}
dx =  \frac{1}{\sqrt{2}}\frac{d\phi}{\sqrt{m^2 U + P}} ,
\ee
and this expression allows, in fact, for an immediate definition of the kink length, namely the distance which is covered by $x$ while $\phi$ runs from $\phi_1$ to $\phi_2$, i.e., 
\be \label{length-int}
L =L(P)=  \frac{1}{\sqrt{2}} \int_{\phi_1}^{\phi_2} \frac{d\phi}{\sqrt{m^2 U + P}}.
\ee
Geometrically, this is the size of the region in physical space where the kink $\phi_{\rm k} (x,P)$ deviates from its vacuum values. Obviously, a kink position or other peculiarities of the solution are not required. 

Now, let us study the case $P=0$ in some more detail. Close to the vacuum $ \phi_1$, i.e., for $\phi = \phi_1 + \delta \phi_1$, the potential will behave like $U \sim c_1 (\delta \phi_1)^{\gamma_1} $ (we assume an algebraic approach to the vacuum). This vacuum approach will lead to a finite contribution to the length integral (\ref{length-int}) for $\gamma_1 <2$, i.e., for a less than quadratic approach to the vacuum. Exactly the same argument applies for the other vacuum $\phi_2$. So a kink at zero pressure has finite length only provided that the corresponding potential approaches both vacua less than quadratically. Kinks with this property are known as compactons \cite{rosenau} - \cite{bazeia}, because the field reaches its vacuum value at a finite distance, i.e., deviates from the vacuum only on a compact domain. Well-known models like the $\phi^4$ model with its standard quadratic approach, on the other hand, lead to an exponential soliton tail and, therefore, to an infinite kink length according to formula (\ref{length-int}). 
As these kinks with exponential tails still have most of their energy and charge concentrated within a finite distance, this result of an infinite length may appear somewhat counterintuitive.  

To better understand the implications of this length definition, let us further study the general case $P\ge 0$. First of all, the same line of reasoning like above shows that, for nonzero pressure, the kink length (\ref{length-int}) is always finite. Next, the energy of a kink with pressure $P$ may be calculated easily,
\be \label{kink-E}
E = E(P)= \int dx (2m^2 U + P) = \frac{1}{\sqrt{2}}\int_{\phi_1}^{\phi_2}  d\phi \frac{2m^2 U + P}{\sqrt{m^2 U + P}} .
\ee
Now we rename $L(P)$ as $V(P)$ (the one-dimensional or generalized ``volume") to be closer to the standard notation of fluid dynamics. Then it is easy to demonstrate that $E$, $V$ and $P$ obey the standard relation \cite{term}
\be
P =-\frac{dE}{dV}
\ee
known from fluid dynamics and (zero temperature) thermodynamics. Indeed,
\be
\frac{dV}{dP} = -\frac{1}{2}\int_{\phi_1}^{\phi_2} \frac{d\phi}{\sqrt{2}} (m^2 U +P)^{-\frac{3}{2}}
\ee
and
\be
\frac{dE}{dP} = \frac{P}{2}\int_{\phi_1}^{\phi_2} \frac{d\phi}{\sqrt{2}} (m^2 U +P)^{-\frac{3}{2}}
\ee
such that
\be
\frac{dE}{dV} = \frac{(dE/dP)}{(dV/dP)} =-P.
\ee
The geometric ``volume" (length) introduced in Eq. (\ref{length-int}) is, therefore, completely equivalent to the thermodynamical or fluid dynamical volume, and the field theory model may be interpreted in thermodynamical and fluid-mechanical terms. 

It follows from Eq. (\ref{kink-eq}) that for nonzero pressure the first derivative of the kink is nonzero at the boundaries, $\phi'_{\rm k} (x_1,P)=\phi'_{\rm k} (x_2, P)=\sqrt{2P}$ (where $\phi_{\rm k} (x_1,P)=\phi_1$,  $\phi_{\rm k} (x_2,P)=\phi_2$). If we want to connect to the vacuum outside the kink (to maintain finite total energy) the first derivative of $\phi_{\rm k} (x,P)$ must, therefore, jump at $x=x_1,x_2$. That is to say, the kink is a continuous function with a discontinuous first derivative. This has to be expected from a physics point of view. The kink $\phi_{\rm k} (x,P)$ is a solution under pressure, and it must be kept stable by applying the same amount of external pressure at its boundaries. The actuation of external pressure at a kink boundary ($x_2$, say) can be avoided if the potential has a third vacuum at $\phi_3 >\phi_2$. Then the upper boundary of the first kink, $x_2$ is, at the same time, the lower boundary of a second kink interpolating between $\phi_2$ and $\phi_3$, with the same pressure, and the two boundary pressures at $x_2$ balance each other. For a potential with $n+1$ vacua we can, thus, construct a chain of $n$ kinks with pressure $P>0$, and external pressure must be applied only at the pre-images of the lowest and the highest vacuum. For potentials with infinitely many vacua we may have infinite kink chains without the necessity of external pressure. Such infinite kink chains might be good models for one-dimensional fluids. We shall study one particular example in the next section.

\subsection{The sine-Gordon kink chain}

The well-known sine-Gordon model with potential 
\be
U_{\rm SG}=1-\cos \phi = 2\sin^2 \frac{\phi}{2}
\ee
provides a simple example of a potential with infinitely many vacua which, in addition, is periodic (invariant under the shift $\phi \to \phi + 2\pi n$). The constant pressure equation (\ref{const-P-eq}) in this case reads
\be
\phi'^2 =4m^2 \left( \sin^2 \frac{\phi}{2} + \tilde P\right) \; , \quad \tilde P \equiv \frac{P}{2m^2}
\ee
or, after the field transformation $\psi = \sin (\phi/2)$,
\be
\psi'^2 = m^2 (1-\psi^2)(\psi^2 + \tilde P).
\ee
This may be brought into the standard form of the differential equation 
\be
\left( \frac{d\psi}{dy}\right) ^2 = (1-\psi)^2 (k^2 \psi^2 +1-k^2)
\ee
for the Jacobi elliptic function ${\rm cn}(k,y)$ by the change of variable
\be
y=\frac{m}{k_P} x \; , \quad k_P \equiv \frac{1}{\sqrt{1+\tilde P}}.
\ee
The kink solution therefore reads $\phi_{\rm k} (x,P) = 2\arcsin ({\rm cn}(k_P,y-y_0))$ or
\be
\phi_{\rm k} (x,P) = 2\arcsin \left[ {\rm cn}\left(k_P,\frac{m(x-x_0)}{k_p}\right) \right] 
\ee
where $x_0$ is an integration constant. ${\rm cn}(k,y)$ is a periodic function with period length $4K_k$ where
\be
K_k \equiv \int_0^\frac{\pi}{2} \frac{d\theta}{\sqrt{1-k^2 \sin^2 \theta}}
\ee
is the complete elliptic integral of the first kind, so one direct way to find the kink length is to use that it must be equal to one-half of the period length, i.e., $2K_k$ in the $y$ variable, which means in terms of the $x$ variable
\be
L=2\frac{k_P}{m} K_{k_P}.
\ee
The same result follows from our general length formula (\ref{length-int}). Indeed,
\be
L=\frac{1}{2m} \int_0^{2\pi} \frac{d\phi}{\sqrt{\sin^2 (\phi/2) + \tilde P}} = \frac{2}{m}\int_0^\frac{\pi}{2} \frac{d\chi}{\sqrt{\sin^2 \chi + \tilde P}}
\ee
where we replaced $\int_0^{2\pi} d\phi$ by $2\int_0^\pi d\phi$ and introduced $\chi = (\phi/2)$. Using the definition of $k_P$ in terms of $\tilde P$ and extracting a factor $k_P$ from the integrand, we further get (introducing $\theta = (\pi/2) - \chi$)
\be
L= 2\frac{k_P}{m} \int_0^\frac{\pi}{2} \frac{d\chi}{\sqrt{1-k_P^2\cos^2 \chi}} =  2\frac{k_P}{m} \int_0^\frac{\pi}{2} \frac{d\theta}{\sqrt{1-k_P^2\sin^2 \theta}}
\ee
which is the desired result. In the same fashion, we may calculate the energy of a kink from Eq. (\ref{kink-E}) with the final result
\be
E = \frac{4m}{k_P} \left( 2E_{k_P} -(1-k_P^2) K_{k_P} \right)
\ee
where
\be
E_k\equiv \int_0^\frac{\pi}{2} d\theta \sqrt{1-k^2 \sin^2\theta}
\ee
is the complete elliptic integral of the second kind. 

We want to emphasize that the periodic sine-Gordon kink chain solution is a known result \cite{mckean} (our presentation follows closely \cite{man-sut-chap5}). The conceptual difference is that the kink chain in \cite{mckean} was found by imposing periodic boundary conditions on the kink, while we imposed a nonzero pressure, instead. Periodic boundary conditions only make sense for periodic potentials (like the sine-Gordon potential), whereas the constant pressure equation (\ref{const-P-eq}) is well-defined for arbitrary  potentials and, in the case of potentials with infinitely many vacua, leads to (in general, non-periodic) infinite kink chains. A simple example might be given by a non-periodic deformation of the sine-Gordon potential like, e.g., 
\be
U=2\sin^2 \left( \frac{\phi}{2}\, \frac{a^2 + \phi^2}{b^2+\phi^2}\right)
\ee
where $a$ and $b$ are nonzero real constants.

\section{General case}
In the preceding section, we defined the geometric "volume" (length) of kinks and found that this geometric volume is, at the same time, the usual fluid-mechanical or thermodynamical volume. In this section, we want to investigate whether this result can be generalized to field theories supporting solitons in higher dimensions. One first important difference is that, in general, higher-dimensional field theories will not be of the perfect fluid type. We shall restrict to field theories with lagrangian densities which are no more than quadratic in first time derivatives (such that a standard hamiltonian exists) and which have no term linear in time derivatives, i.e., no Chern-Simons like terms (this last restriction is only introduced to simplify the discussion and probably can be relaxed without problem). Then the energy-momentum tensor for static configurations has the non-zero components $T_{00} = \mathcal{E}$ and $T_{ij}$, while $T_{0i}=0$. For a perfect fluid, the space-space part of the energy-momentum tensor for static fields reads $T_{ij}=\delta_{ij}\mathcal{P}$, where $\mathcal{P}$ is the pressure density. 
In this case, energy-momentum conservation still implies that the pressure density for static soliton solutions must be constant, $\mathcal{P} =P=$ const.
There exist field theories leading to perfect fluids \cite{term}, but they are the exception rather than the rule. For general field theories we, therefore, need a definition for an average pressure before we can try to find a related thermodynamical volume. We shall use the standard definition
\be
\mathcal{P} = d^{-1}{\rm tr} {\bf T} \equiv d^{-1} \sum_i T_{ii}
\ee
for the average pressure density, and for the average pressure
\be
 P = V^{-1} \int d^d x \mathcal{P}.
\ee
Here, $V$ is the geometrical volume of the soliton, i.e, the volume of the region of $\mathbb{R}^d$ where $\mathcal{E}$ is different from zero. The statement now is that, under some additional assumptions on the underlying field theory, the geometrical volume is, at the same time, also the thermodynamical (fluid-mechanical) volume. First we need some notation. The fields $\phi^a$, $a=1, \ldots,m$ of the field theory take values in a certain target space manifold $\mathcal{M}$, and static field configurations $\phi^a (\vec x)$ belong to the space of maps $\{\boldsymbol{\Phi}\}$ where
\be
\boldsymbol{\Phi}: \mathbb{R}^d \to \mathcal{M}:\; \vec x \to \phi^a (\vec x).
\ee
We assume, of course, that the field theory in question leads to static soliton solutions, at all.
Further, we assume that the field theory has at most a discrete set of vacuum configurations $\phi^a_{n,{\rm vac}}$, $n=1, 2, \ldots$, one of which the field must approach in the limit of infinite $|\vec x |$ for finite energy configurations (in the case of one space dimension, the field may approach two different values in the two directions $x\to \pm \infty$, because these two directions cannot be continuously rotated into each other). This may occur because the energy density contains a potential with a discrete set of vacua. If the vacuum of the potential is not unique (e.g., a Higgs-type potential), or if there is no potential, it may still hold that for finite energy configurations the field must approach the same field configuration $\lim_{|\vec x| \to \infty} \phi^a (\vec x)= \phi_{\rm inf}^a$ in all directions. If the field theory has some internal symmetries, then these internal symmetries might transform $\phi^a_{\rm inf}$ nontrivially. In this case, we require that the condition of finite energy imposes that the symmetry must be broken spontaneously, i.e., one of the elements of the symmetry orbit space of $\phi^a_{\rm inf}$ must be chosen as the physical vacuum $\phi^a_{\rm vac}$. The assumption of a discrete set of vacua $\phi^a_{n,{\rm vac}}$, or of a unique vacuum configuration $\phi^a_{\rm vac}$ after spontaneous symmetry breaking, is a nontrivial assumption. 
It is imposed in order to guarantee that vacuum regions in $\mathbb{R}^d$ (i.e., regions where the field takes one of its vacuum values) are, at the same time, regions of zero energy density, except for regions of measure zero (lower-dimensional submanifolds, e.g., boundaries).
If we allowed for vacuum manifolds with more than zero dimensions, then gradients along these vacuum manifolds could produce a nonzero energy density. Our assumption still covers a lot of models, mainly based on scalar fields, like the real scalar field models in $1+1$ dimensions of the last section, nonlinear sigma models, Skyrme-type models in arbitrary dimensions \cite{skyrme}, \cite{skyrmions}-\cite{baby-sk}, or the Faddeev-Hopf model of knot solitons \cite{FN}, \cite{ferr}. After a slight modification of our considerations, it probably also covers many non-relativistic soliton models, although the explicit formulation in this letter is for relativistic (Poincare-invariant) field theories.
  
On the other hand, the above assumption does not cover cases like the vortices of the Abelian Higgs model or the t'Hooft-Polyakov monopole. In these cases, non-trivial gradients along the vacuum manifold exist but are exactly compensated by gauge fields. We do not exclude the possibility that 
 the arguments given below may be generalized to models where the above assumption does not hold, such that cases like monopoles and vortices may be covered. This generalization, however, is beyond the scope of the present letter \footnote{In the case of vortices, it is possible to choose a singular gauge such that the
scalar field of the vortex has a direction independent limit at infinity, and the vacuum manifold is effectively reduced to a point, similarly to the case of spontaneous symmetry breaking. To include this case into our considerations, however, still requires an adequate treatment of the singularity and an appropriate inclusion of the gauge field in our generalized step function (\ref{gen-step}). We thank the referee for pointing out this possibility.}.

Next, we define the following generalized step function on target space
\be \label{gen-step}
 \Theta (\phi^a) =
\left\{
\begin{array}{c}
1 \quad \mbox{for} \quad \phi^a\notin  \mathcal{V} \\
0 \quad \mbox{for} \quad \phi^a \in \mathcal{V}
\end{array}
\right.\, , \qquad
\mathcal{V} =\{ \phi^a_{n,{\rm vac}} \}
\ee
where $\mathcal{V}$ is the (zero-dimensional) vacuum manifold.
Further, we define the locus function of a static field configuration $\phi^a (\vec x)$ as the pullback $\boldsymbol{\Phi}^* ( \Theta (\phi^a))$ of $ \Theta (\phi^a)$ under $\phi^a (\vec x)$, and the locus set of the static field $\phi^a (\vec x)$,
\be
\Omega = \{ \vec x \in \mathbb{R}^d \; | \;  \boldsymbol{\Phi}^* ( \Theta (\phi^a)) \equiv \Theta (\phi^a (\vec x))=1 \},
\ee
i.e., the set $\Omega \subset \mathbb{R}^d$ where the static field configuration is located (deviates from the vacuum).
Now we introduce the extended static energy functional 
\be \label{en-func}
E_{\bf e} (V,P)=\int d^dx \; \mathcal{E} [\phi^a] + P\left( \int d^d x  \Theta (\phi^a (\vec x))-V \right)  , 
\ee
where $\mathcal{E}$ is the energy density and $P$ is a Lagrange multiplier imposing the condition that all possible solutions of the variational problem (\ref{en-func}) must have  geometric volume $V$, i.e., $\int d^d x  \, \Theta (\phi^a (\vec x)) = \int_\Omega d^d x =V$. 
Obviously, $P$ obeys the thermodynamical relation
\be
\left( \frac{\partial E}{\partial V} \right) =-P, \label{p1}
\ee
 by construction. 
We still have to show that the Lagrange multiplier $P$ is, indeed, the average pressure as defined above. 
To prove it,  we will adapt some methods developed in \cite{sp1} to our case. Concretely, we act with a scaling transformation $x^i \rightarrow e^\lambda x^i \sim (1+\lambda)x^i$ on the fields $\phi^a (\vec x)$ in the extended energy functional above. Then, we use the fact that both functionals above may be generalized to metric-dependent ``general-relativistic" functionals which are invariant under general coordinate transformations and, in particular, under linear coordinate transformations $\boldsymbol{x} \to \Lambda \boldsymbol{x}$, i.e., 
\be
E[\phi^a (\boldsymbol{x})] \equiv \int d^d x \mathcal{E}[\phi^a(\boldsymbol{x})] \; \Rightarrow \;  E[g_{ij},\phi^a (\boldsymbol{x})]
\equiv \int d^d x \sqrt{-g} \mathcal{E}[g_{ij},\phi^a(\boldsymbol{x})] 
\ee
and
\be
\int d^d x \Theta (\phi^a (\boldsymbol{x})) \; \Rightarrow \; \int d^d x \sqrt{-g}\Theta (\phi^a (\boldsymbol{x}))
\ee
where $g_{ij}$ is a general metric in $\mathbb{R}^d$ and $g={\rm det} g_{ij}$. Coordinate invariance just means
\be
E[g_{ij},\phi^a (\boldsymbol{x})] = E[(\Lambda^* g)_{ij},\phi^a (\Lambda \boldsymbol{x})]
\ee
and the same for the second functional. Functionals which allow for this coordinate-invariant ``general-relativistic" generalization are called ``geometrically natural" in \cite{sp1}. But this invariance implies that
\be
E[g_{ij},\phi^a (\Lambda \boldsymbol{x})] = E[((\Lambda^{-1})^* g)_{ij},\phi^a ( \boldsymbol{x})] ,
\ee
so transformations on $x^i$ may be transferred to transformations on $g_{ij}$, which will simplify our task. Concretely, for the scaling transformation we have $\Lambda^i{}_j =(1+\lambda)\delta^i{}_j$ and $((\Lambda^{-1})^*)_i{}^j =(1-\lambda)\delta_i{}^j$ and, therefore, up to order $\lambda$
\be
((\Lambda^{-1})^* g)_{ij} =(1-2\lambda )g_{ij} \quad \Rightarrow \quad \delta g_{ij} = -2\lambda g_{ij} = +2\lambda \delta_{ij}
\ee
(remember we use the ``mostly minus" metric convention $g_{ij} = -\delta_{ij}$).
Now we need to know how the energy functional changes under a variation of the metric. For our metric conventions, the $d+1$ dimensional matter action $S$ is related to the energy-momentum tensor via
\be
\frac{\delta S}{\delta g^{(d+1)}_{\mu\nu}(x^\rho)} = -\frac{1}{2}\sqrt{-g^{(d+1)}}T^{\mu\nu}(x^\rho).
\ee
But for the class of theories we consider, for static fields it holds that $\mathcal{E} = -\mathcal{L}$, which implies
\be
\frac{\delta E}{\delta g_{ij}(\vec x)} = +\frac{1}{2}\sqrt{-g}T^{ij}(\vec x).
\ee
For the variation of the energy functional we then find
\be
\delta E = \left. \int d^d x \, \delta g_{ij}(\vec x) \, \frac{\delta E}{\delta g_{ij}(\vec x)}\right|_{g_{ij}=-\delta_{ij}}  = \int d^d x \, 2\lambda \delta_{ij} \, \frac{1}{2} T^{ij} = \lambda \int d^d x \sum_i T_{ii}.
\ee
For the second functional we just need the obvious result $\delta \sqrt{-g} = -\lambda d\sqrt{-g}$. 
Hence, for the extended energy functional (\ref{en-func}) we get, off-shell and in first order in $\lambda$, 
\be
\delta E_{\bf e} = \lambda \left( \int \sum_i T_{ii}d^d x -d\, P\int d^dx \Theta (\phi^a (\vec x)) \right).
\ee
But on-shell, i.e., for a particular soliton solution, this must be zero because any solution of the (extended) Euler-Lagrange equations is a stationary point and, therefore, 
\be \label{PV}
P=\frac{d^{-1} \int_\Omega \sum_i T_{ii}d^d x}{\int_\Omega d^d x} =V^{-1} \int_\Omega d^d x \mathcal{P}
\ee 
which is what we wanted to prove. 

We remark that, unless the field theory is equivalent to a perfect fluid, different soliton solutions will, in general, lead to different (average) pressures $P$, even for the same volume $V$. Different solutions for the same volume are guaranteed to lead to the same pressure only if they are related by symmetry transformations (although it may happen accidentally that some solutions not related by symmetry still have the same pressure). We further remark that it follows from (\ref{PV}) that, for nonzero pressure $P$, the  volume $V$ is always finite (here we assume that $\mathcal{P}$ has no singularities). For zero pressure, there are several possibilities. The first is that the volume is infinite (solitons with infinite tails). 
Turning on a nonzero pressure makes the volume jump from an infinite to a finite value. This implies that in this case the compressibility at zero pressure is infinite. The second possibility is that the pressure density $\mathcal{P}$ is identically zero (which certainly is the case for perfect fluids). The volume may then be either finite (compactons) or infinite (solitons with infinite tails). The third possibility seems to be that the pressure density, while not being identically zero, averages to zero. We are not aware of a specific example where this happens, and we do not know whether this possibility really exists (or whether, even for theories which are not perfect fluids, zero pressure plus finite volume imply zero pressure density). Finally, we remark that, for the particular case of the Skyrme model, similar considerations have already been briefly introduced in \cite{squeeze}, although without an explicit proof. Here, we added the proof and gave a more detailed discussion, and we generalized to a larger class of field theories in arbitrary dimensions.  

\section{Discussion}
The volume definition proposed and developed in the present letter was mainly motivated by its mathematical properties (the equivalence of geometrical and thermodynamical volume), so the physical meaning and interpretation of this volume may depend both on the particular solitonic field theory under consideration and on the physical system it is supposed to describe. 
As a first example, let us consider a field theory supporting solitons with compact support (compactons). In this case, together with the elementary compacton (the solution with minimal energy), there exists $n$-compacton solutions consisting of $n$ non-overlapping elementary compactons surrounded by empty space (vacuum). The energy and geometric volume of a $n$-compacton solution are, then, $n$ times the energy (geometric volume) of the elementary compacton. In this system, it is plausible to interpret the compactons as particles, such that the geometric volume equals the total volume of all particles of a given solution. Further, the $n$-compacton solution may be enclosed in a finite box without any cost in energy provided the box is sufficiently large to enclose all compactons. This $n$-compacton solution in a finite box, therefore, resembles a classical gas of $n$ particles at zero temperature, and the box volume may be identified with the gas volume (a specific example of such a field theory was discussed, e.g., in \cite{term}). For nonzero pressure, on the other hand, the individual compactons are compressed and move towards each other, empty space is expelled, and the resulting system is then either a fluid or a solid (a crystal), depending on the amount of symmetry of the underlying static energy functional. The geometric volume should be identified with the volume of the fluid or the solid in this case.   

Next, we want to consider a field theory giving rise to topological solitons with (e.g. exponential) tails, with infinite geometrical volume, and where field configurations may be classified by an (integer-valued) topological degree interpretable as particle number. Furhter, we assume that solutions with particle number $n>1$ may be interpreted as collections of loosely bound $n=1$ solitons, where small regions of rather large energy density (the soliton positions) are surrounded by large regions of very small but nonzero energy density with field values close to (but different from) the vacuum. Specific examples of field theories with these characteristics are the "loosely bound" Skyrme model \cite{loosely} and its lower-dimensional baby version ("aloof baby skyrmions") \cite{aloof}. Due to the small binding, it is still reasonable to interpret this system as a system of weakly interacting particles at zero temperature. Concretely, as solutions with a given particle number (baryon number) come in discrete sets (one global minimum, and several almost degenerate local minima) with a fixed particle configuration for each solution, they are best interpreted as soft (weakly bound) crystals or lattices. It is, then, no longer plausible to identify the (in this case, infinite) geometric soliton volume with the total particle volume. Indeed, the zone of small field values between the local energy density maxima is better interpreted as an interaction zone mediating the forces between the individual elementary solitons (particles). 
A more physical interpretation may, then, be like follows. Individual particle sizes (volumes) are defined as the volumes of small balls about the local energy density maxima (the positions of the elementary solitons). Here, e.g., the generalized RMS radii mentioned in the introduction may  
be used to calculate these particle volumes. These definitions are not unique, so the specific definition chosen should be adapted to the model under study and to the physical system it is supposed to describe. The geometric (and, at the same time, thermodynamical) volume, on the other hand, should be identified with the volume of the whole system, i.e., with the volume of the soft lattice (crystal) of interacting particles, which covers both the particles and the interaction zone. This interpretation receives further support if one considers what happens when a small but nonzero pressure is introduced. The geometric volume is then finite, resulting in a lattice of weakly interacting particles in a finite box at nonzero pressure. 

It should be emphasized that the physical interpretations of the geometrical volume in the two examples above are related to a particle-like interpretation of their solitonic solutions. Concretely, we not only interpreted the elementary solitons as (elementary) particles, but also the higher solitons as collections of (bound or unbound) particles. This latter interpretation, however, is not always plausible. In general, a soliton describes both a "particle" and its interactions, and a clear separation may not be possible. That is to say, some field theories have higher soliton solutions where elementary soliton substructures cannot be identified, because either the interactions are so strong that the elementary solitons cluster into new substructures, or the elementary solitons get completely dissolved,  and the higher soliton should just be interpreted as a collective, coherent excitation of the underlying basic scalar fields, without a clear relation to smaller solitonic constituents or to a particle picture. The physical interpretation of the geometric volume will be different in these cases, whereas its mathematical characteristics remain, of course, unchanged. 

\section{Conclusions}
We demonstrated that, for a large class of nonlinear field theories supporting soliton solutions, there exists an unique definition of the corresponding soliton volume which is equal to the geometrical volume and, at the same time, to the thermodynamical (fluid-dynamical) volume.
In a first instant, this may appear mainly as a peculiar observation, but we think that our considerations, in addition to providing this volume definition, demonstrate a close relationship between nonlinear field theories with solitons, on the one hand, and concepts of fluid dynamics and thermodynamics, on the other hand, and that this relationship will be productive both for a deeper analysis of nonlinear field theories and for the recent attempts to find field theoretic formulations of fluid dynamics and hydrodynamics \cite{field-fluid}. 
Indeed, depending on the structure of their static soliton solutions, some nonlinear field theories may lead to rather nontrivial and interesting zero temperature thermodynamics. If there exists only one solution - up to symmetries - then this solution will lead to a unique equation of state (EoS) $V(P)$. On the other hand, if there exists a discrete set of different solutions (up to symmetries) in a given topological sector (i.e. deformable into each other), such that one solution is a true minimum whereas the other ones are local minima, then this may give rise to phase transitions. That is to say, the different solutions will generically lead to different EoS $V_i(P)$ and to different energy relations $E_i(P)$, and it might happen that at a certain value $P=P_0$ of the pressure one of the local energy minima at $P=0$ turn into the new global minimum, such that a first-order phase transition  occurs. In particular, for theories where a particle interpretation of higher soliton solutions is possible (as discussed in the previous section), and where the space-time symmetries are just the Poincare symmetries, the discrete set of different solutions might, e.g., correspond to different crystal structures, and the phase transitions to transitions between them.
  
If there exist whole families of solutions, degenerate in energy at zero pressure and distinguished by some continuous parameters (beyond the ones parametrizing the symmetries) then, a priori, there seem to exist two possibilities. Either all solutions within this parameter family lead to the same energy and the same EoS also at nonzero pressure, or different solutions produce different energies and/or different EoS. If this second possibility is realized, then the solution parameters entering the energy expression and/or the EoS should play the role of further thermodynamical variables. 
For topological solitons which support a topological charge, the corresponding chemical potential may also be easily introduced, see \cite{squeeze} in the case of skyrmions.

Finally, we also provided a rather detailed discussion of the exemplary case of kinks in $1+1$ dimensions, including a generalized method for the construction of infinite kink chains, which might be of some independent interest.

\section*{Acknowledgement}
The authors acknowledge financial support from the Ministry of Education, Culture, and Sports, Spain (Grant No. FPA 2014-58-293-C2-1-P), the Xunta de Galicia (Grant No. INCITE09.296.035PR and Conselleria de Educacion), the Spanish Consolider-Ingenio 2010 Programme CPAN (CSD2007-00042), and FEDER. AW thanks S. Krusch for discussion and D. Harland for inspiring remarks on the average pressure issue. Further, the authors thank J. Sanchez-Guillen for helpful comments.

\end{document}